# Designing an Improved Deep Learning-based Model for COVID-19 Recognition in Chest X-ray Images: A Knowledge Distillation Approach

AmirReza BabaAhmadi*

*School of Mechanical Engineering, College of Engineering, University of Tehran, Tehran, Iran (email: babaahmadi.amir@ut.ac.ir)*

Sahar Khalafi

*Department of Electrical and Computer Engineering, Isfahan University of Technology, Isfahan, Iran*

Masoud ShariatPanahi

*School of Mechanical Engineering, College of Engineering, University of Tehran, Tehran, Iran*

Moosa Ayati

*School of Mechanical Engineering, College of Engineering, University of Tehran, Tehran, Iran*

## Abstract

**Context:** COVID-19 has had a significant impact on society, leading to the need for accurate identification and appropriate medical treatment. Various manual and automatic feature extraction techniques have been explored to address this issue. However, the computational requirements of deep learning models hinder their accessibility, especially for institutions and societies with limited resources.

**Objectives:** This study aimed to address two primary goals: reducing computational costs for running the model on embedded devices, mobile devices, and conventional computers, and improving the model's performance compared to existing methods. The objective was to ensure high performance and accuracy for the medical recognition task.

**Methods:** Two neural networks, VGG19 and ResNet50V2, were employed to enhance feature extraction from the dataset. These networks provided semantic features, and their feature vectors were merged in a fully connected classifier layer to achieve satisfactory classification results for normal and COVID-19 cases. To overcome the computational burden, MobileNetV2, a network known for its efficiency on mobile and embedded devices, was adopted. Knowledge distillation was applied to transfer knowledge from the teacher network (ResNet50V2 and VGG19) to the student network (MobileNetV2), improving its performance for COVID-19 identification.

**Results:** Pre-trained networks and 5-fold cross-validation were employed to evaluate the proposed method. The model achieved an accuracy of 98.8% and an F1 score of 99.1% in detecting infectious and normal cases.

**Implications:** The results demonstrate the superior performance of the proposed method. The student model, with acceptable accuracy and F1-score using cross-validation, is suitable for conventional computers, embedded systems, and clinical experts' cell phones. This approach provides a cost-effective solution for COVID-19 identification, enabling wider accessibility and accurate diagnosis. Compared to previous works, our method demonstrates superior results in terms of accuracy, F1 score, and other related metrics, with an improvement of at least 1%.

## 1. Background and Objectives

COVID-19 has been a significant threat to human life and health in recent years. It has had a detrimental effect on healthcare systems worldwide. Unfortunately, COVID-19 is highly contagious and transmissible. As a result, it is inevitable to develop prediction systems capable of rapidly diagnosing it and averting its adverse effects. Numerous scientists have conducted multiple studies to develop human-level prediction and diagnosis systems to aid communities in combating this disease.

These methods are typically effective at detecting COVID-19 using CT and chest X-ray images. However, the fatal flaw is that automatic feature extraction techniques, particularly deep learning models, require a significant amount of computation to perform a specific task. Numerous hospitals and institutions are unable to procure expensive hardware systems capable of performing these algorithms.

Additionally, in the absence of medical experts, the health of the population in many developing countries may be jeopardized. In some developing countries, the scarcity of human experts capable of identifying medical diseases from medical images is a stark reality. As a result, an effort is made to offer an alternate solution to remedy this issue in this study.

Recognizing COVID-19 is the first step toward treatment. The main objective of this paper is an attempt to address this step. COVID-19 is typically accompanied by symptoms including coughing, colds, and shortness of breath, among others. According to the WHO, respiratory problems are the fatal symptoms of COVID-19 and can be identified using CT scans or chest X-rays obtained in hospitals and medical clinics.

Deep learning and its associated techniques have been widely applied to computer vision tasks such as medical image analysis in recent years. Examples include tumor detection [1], diabetic retinopathy [2], among others.

Considering that medical image analysis is one of the most exciting applications of computer vision and deep learning, automatic feature extraction algorithms can be critical in diagnosing diseases at levels comparable to human experts. Due to the critical nature of the initial diagnosis, researchers have identified the accuracy of this step as a significant challenge.

## *1.2 Related Works*

The following section examines related works and studies to create a solid framework for our research. In [3], a model based on stacked convolutional neural networks with multiple pre-trained networks was proposed; however, this work suffers from high computational costs. In [4], a residual-layer-based neural network with a modified kernel was used to predict the presence of COVID-19 in normal and pneumonia chest X-ray images. This work addressed the issues of vanishing gradient and degradation; however, their modified kernel did not significantly reduce the computational costs and is not feasible for use on edge devices.

In [5], a model based on an ensemble learning classifier is proposed to address the issue of an imbalanced dataset. However, using ensemble models has several disadvantages, such as reduced interpretability, increased deployment complexity, and higher computational requirements.

Another paper [6] introduced a combination of LSTM (for detection) and a convolutional neural network (for recognition). This method achieved a 99.4% accuracy metric. However, LSTM is primarily designed to capture temporal dependencies in sequential data, and it may not fully exploit the spatial relationships crucial in medical image processing. Additionally, this method lacks interpretability.

In [7], a model using LSTM and GAN architectures is described. This method eliminates the need for an additional network for feature extraction in the binary classification task. However, GANs can suffer from mode collapse, where the generated images are limited in variation or fail to capture the full diversity of the target distribution. In medical image processing, where capturing the full range of anatomical or pathological variations is crucial, mode collapse can lead to unrealistic or biased images, limiting the usefulness of GANs in certain applications. Evaluating and validating GAN-generated medical images can also be challenging, as traditional evaluation metrics used in computer vision may not fully capture the quality or clinical relevance of the generated images.

In paper [8], the presence of COVID-19 in medical images is investigated using a deep neural network based on a CNN with additional components in different layers. Unfortunately, this work did not focus on computational costs, and its sole objective was to increase accuracy and evaluation metrics.

In [9], CNN and SVM were combined to analyze COVID-19 CXR images. However, using SVM for classification can be time-consuming and cannot handle massive amounts of data. Another drawback of using SVM is that the training process will no longer be end-to-end.

In [10], a new framework based on the use of DarkNet53 to process chest X-ray images is proposed. DarkNet53 was originally designed as a custom backbone for the YOLO network for object detection purposes and may not be suitable for classification tasks.

Paper [11] conducted a comparative study to determine the best-performing pre-trained deep learning architectures. Fifteen models were examined to identify the optimal architecture, and according to this study, VGG19 was found to be the most accurate pre-trained model.

Another comparative study [12] classified COVID-19 from normal cases and pneumonia using novel pre-trained architectures such as DenseNet201, InceptionV3, ResNet50, and several other networks. However, both of these papers solely focused on comparing the performance of pre-trained networks and did not consider the practical applicability of these networks in hospital settings.

In paper [13], a new model based on GAN and CNN was introduced. Multiple classifiers were used to perform the classification task in work. Furthermore, this paper developed a COVID-19 segmentation model using a unique dataset. However, we believe that using multiple classifiers does not have a remarkable effect on classification tasks and can lead to high computational requirements, making this method impractical for real-world usage.

In [14], pre-trained architectures were compared, and modifications were made to reduce trainable parameters and enable faster algorithm training. This is a valuable contribution; however, the issue of forgetting (catastrophic forgetting) must also be addressed, which has not been discussed in this paper.

In [15], a variety of pre-trained architectures, including AlexNet, VGG16, and the ResNet family, were examined. Additionally, this article focused on evaluating freezing layers and other configurations used in transfer learning methods.

A study was conducted in [16] to determine the best candidate for COVID-19 recognition. According to this research, the Inception Network achieved outstanding results for COVID-19 detection.

Another comparative study was conducted in [17] using DenseNet121, ResNet50, VGG16, and VGG19. These works were suitable for finding suitable backbones for COVID-19 recognition. However, they did not address computational costs and deployment problems. Additionally, they did not contribute to the technical aspects of improving architectures.

In [18], a binary classification task was performed, and fine-tuning was applied to the final layer of pre-trained SqueezeNet, DenseNet121, ResNet50, and ResNet18 networks. However, the problem with this work is that fine-tuning alone does not solve the challenge of high computational resources required to train these networks and their deployment configurations.

Paper [19] presents a voting-based procedure for evaluating predictions from VGG16, InceptionV3, and ResNet50. The final layer compares the performance of all the mentioned networks simultaneously and selects the best result. While this model demonstrates good performance, its deployment for practical applications is complex and requires significant modifications to reduce computational expenses.

In [20], an assessment of the effect of preprocessing on the results of CNN-based networks is conducted. However, pure preprocessing only leads to performance improvement and does not address the computational burdens associated with these networks.

In [21], the development of a customized network called COVID-Net for CXR and chest X-ray medical images is described. However, COVID-Net suffers from complexity, and the deployment process for COVID-Net requires substantial effort to reduce computational burdens.

Another study [22]compares the classification results of COVID-19, bacterial pneumonia, and viral pneumonia using state-of-the-art algorithms and CNN architecture. While interesting, this study still utilizes pre-trained networks with minimal modifications, and the aforementioned problems remain unresolved.

In yet another study [23], PCA (Principal Component Analysis) was used to improve the efficiency of feature extraction. However, we believe that PCA suffers from linearity and sensitivity to outliers, making it less ideal for the task at hand.

In [24], YOLO Networks with additional convolution layers involving modified kernels were employed to detect COVID-19 from chest X-ray images. As previously mentioned, YOLO with modified kernels is well-suited for object detection tasks, but it is not specifically optimized for image classification, which may introduce problems in classification tasks.

In [25], the performance of the Efficient-Net family in detecting COVID-19 and pneumonia in normal cases was examined. This network is well-suited for practical applications. However, this work did not focus on addressing the mitigation problem in real-world practical applications. The training was primarily conducted in academic settings.

In [26], a network called CoroNet was introduced, based on the Xception network, to perform multi-class classification including pneumonia-viral, pneumonia-bacterial, and COVID-19 categories. This architecture is lightweight, but its accuracy and other metrics are not promising enough to be considered an eligible candidate for real-world applications.

There are also related works that employ a fuzzy-enhanced deep learning approach for COVID-19 recognition in chest X-ray images. The authors claim that their algorithm improves performance for imbalanced datasets and portable chest X-ray images [27].

In another study, a novel neuro-evolution algorithm has been proposed for COVID-19 recognition in chest X-ray images. The researchers utilized a K-nearest neighbor (KNN) algorithm to classify infected cases from normal cases [28]. However, we believe that KNN is unable to handle massive amounts of data and is not an ideal choice as a classifier for modern deep learning algorithms.

All the papers we examined did not address the issue of mitigating problem and reducing computational expenses simultaneously. Hence, our focus was on developing a method that tackles both aspects, as elaborated in the remainder of this paper.

To overcome these challenges, we explore the application of the knowledge distillation technique, also known as the teacher-student model. Our primary objectives are twofold: firstly, to achieve high accuracy in detecting COVID-19 in chest X-ray images, and secondly, to significantly reduce the computational power required during training. This reduction in computational requirements enables the utilization of conventional computers that are accessible in deprived countries or cities. Additionally, we aim to leverage the learned model on low-power chip-based mobile devices for efficient inference in disease recognition from medical images.

To accomplish these objectives, we propose a novel architecture that leverages dual networks to extract richer semantic features. We transfer this knowledge to a lightweight network, facilitating efficient embedding within software engineering and production pipelines. To construct our two-class dataset, we have concatenated two publicly available datasets from Kaggle, as elaborated in Section 4.

Our study yields highly promising results and evaluation metrics. We achieve an impressive accuracy rate of 98.8% and an F1 score of 99.1%, surpassing the performance of previously published works. As a result, our methodology exhibits significant potential for reducing computational power requirements while increasing the accuracy of neural networks specifically tailored for embedded devices.

Key Contributions:

1. Computational Efficiency: By employing knowledge distillation, our approach significantly reduces the computational demands during training, enabling the utilization of conventional computers in deprived regions. It also mitigates forgetting problem which occurs in deep neural networks.

2. Mobile Device Compatibility: Our model is designed to be efficiently deployed on low-power chip-based mobile devices, enabling cost-effective and portable disease recognition in medical images.

3. Enhanced Accuracy: The achieved accuracy of 98.8% and an F1 score of 99.1% demonstrate the superior performance of our proposed methodology, surpassing previously published works.

4. Potential for Embedded Devices: Our methodology holds the potential to enhance the accuracy of neural networks specifically designed for embedded devices, expanding the range of applications in resource-constrained environments.

This paper is structured as follows: The second section provides an overview of previous publications on the task. The third section describes the knowledge distillation method and explains why it was incorporated into the algorithm. Section four discusses the dataset that was used to conduct this research. Section five describes the training phase and the evaluation metrics. Finally, section six presents the evaluation results of the proposed model and compares them with related works, while section seven summarizes the findings of our paper.

## 2. Methods

Nowadays, Deep Neural Networks have revolutionized the field of computer vision. Their applications have been extensively investigated in a variety of fields, including self-driving cars [29], medical image analysis [30], [31], and agriculture [32], among others. CNNs have established themselves as the most effective tools for automatic feature extraction in computer vision, NLP, speech processing, and video classification tasks.

This paper demonstrates how CNN-based neural networks can improve semantic feature extraction for binary classification tasks involving COVID-19 and normal cases. DenseNet [33], ResNet [34], VGG [35], Xception [36], and Mobile-NetV2 [37] are some of the most powerful pre-trained networks. VGG19 is an extension of VGG16, featuring sixteen convolutional layers and three fully connected layers. It also uses five MaxPool layers and a final SoftMax layer.

ResNet50V2 is an enhanced version of ResNet50, outperforming previous versions such as ResNet101. It incorporates several new connections between different blocks, allowing it to achieve high accuracy in the ImageNet competition. MobileNetV2 is a convolutional network designed with depth-wise convolution layers to improve accuracy while reducing computational costs, achieved by decreasing trainable parameters. It is suitable for embedding in mobile phones, tablets, and other conventional embedded devices due to its use of inverted residual blocks.

We have trained and evaluated numerous paired neural networks to select the optimal architectures for constructing the teacher network. Additionally, we have trained several pre-trained neural networks for the student architecture. As a result, we have chosen VGG19 and ResNet50V2 due to their outstanding performance when compared to other architectures. The same rationale holds true for the selection of the student architecture.

Since both ResNet50V2 and VGG19 generate the same size output layer (feature vector), their feature vectors were combined to produce a richer semantic feature set for the specified task. A CNN layer with a kernel size of 1 and 1024 filters was added to this architecture. The network's output was then flattened and streamed into fully connected layers. Notably, no activation function was used in the final CNN layer. The classification layer consisted of 64 neurons in the fully connected layers, with a dropout rate of 0.5. Another fully connected layer was added, serving as the final layer with a neuron count equal to the number of classes in the problem.

The entire architecture is referred to as the teacher network, with MobileNetV2 serving as the student network in a broader context known as the teacher-student model or knowledge distillation framework. The structure of the teacher model is depicted in Figure 1. The feature size of the tensors for both ResNet50V2 and VGG19 is 10*10*2048, and the size of the concatenated feature is 10*10*4096.

In our study, we employed 5-fold cross-validation to select optimal hyperparameters. Cross-validation is a robust technique widely used for hyperparameter tuning and model evaluation. With 5-fold cross-validation, our dataset was divided into five subsets or folds. We iteratively trained our model on four folds while using the remaining fold for validation. This process was repeated five times, each time using a different fold as the validation set. By systematically varying the hyperparameters and evaluating the model's performance across the five folds, we were able to identify the hyperparameter configuration that yielded the best overall performance. The use of 5-fold cross-validation allowed us to effectively assess and fine-tune our model's hyperparameters, ensuring the reliability and generalizability of our findings.

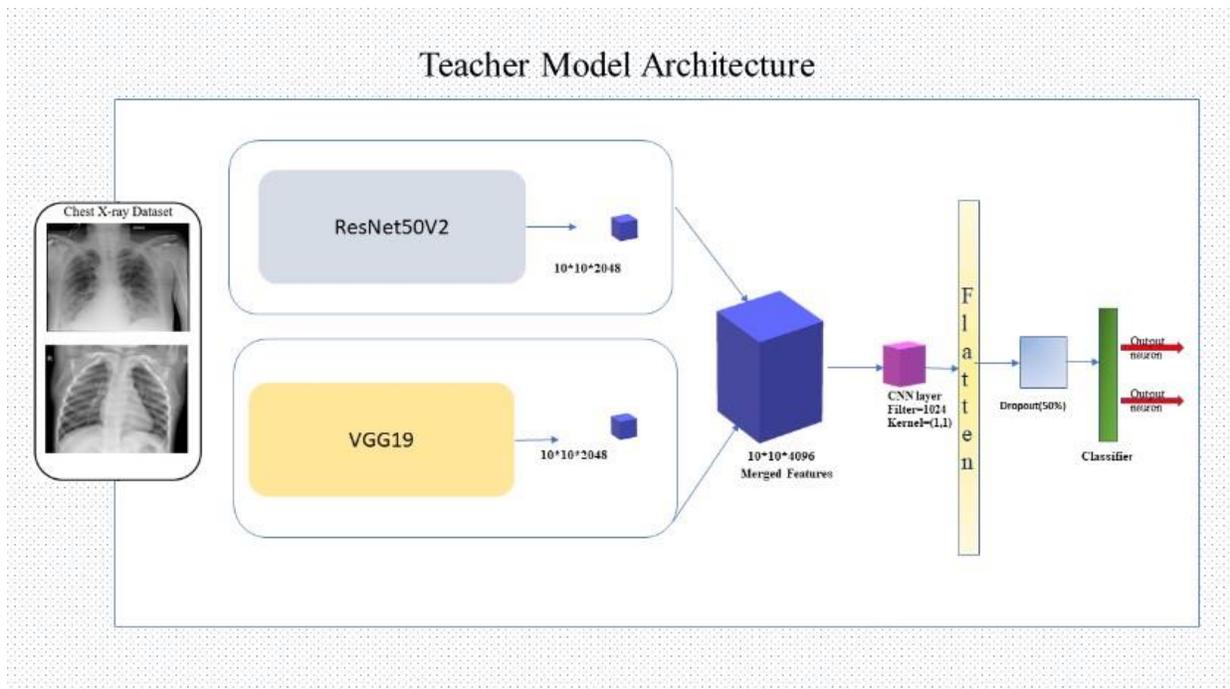

**Fig 1. Teacher Model Architecture**

## 3. Knowledge Distillation

Hinton et al. pioneered Knowledge Distillation in [38]. Knowledge Distillation (KD) is a process that involves training a smaller network to imitate the behavior of a larger network. The purpose of designing a complex network as a teacher is to learn more sophisticated features and achieve better results. However, when running our network on a standard computer or embedded device, we often encounter limitations in memory size and computational cost, leading to frequent issues. Hence, a solution is required to address these concerns.

To distill knowledge from the teacher to the student, a weighted average (mean) is necessary. The initial objective function is Cross-Entropy with soft targets, which is calculated through the softmax function in the smaller network using a higher temperature. Soft targets are generated by a larger architecture or network. The second objective function is Cross-Entropy with valid labels, which utilizes the softmax output from the student model with a temperature of zero.

The teacher network and the student network start receiving training data in parallel. The teacher model incorporates a softmax function with temperature in its output. In contrast, the student model generates two distinct outputs. The first output is softmax with temperature, while the second output consists of the standard softmax. The purpose of the student model is to produce softened probabilities, which correspond to the output of the teacher model. The loss of knowledge distillation is calculated using the following formula:

$$L_{KD} = \alpha KL(p.q)T^2 + (1-\alpha)L(W_s.x) \qquad (1)$$

Here, $p$ and $q$ denote probabilities generated by the student and teacher networks in a specific temperature ($T$), respectively, and $KL$ denotes the Kullback-Leibler divergence, that measures the level of distinction between two probabilistic distributions. The Cross-Entropy of the student model with $T=1$ is ($LW_s.x$). According to [38], $\alpha$ and $T$ are hyperparameters where the greater the value of $\alpha$, the better the learning experience for the student model.

During the distillation phase, back-propagation should only be performed in the student network, as the teacher network has already fine-tuned its parameters. The transfer of knowledge from the teacher to the student model takes place throughout the distillation procedure. It is worth noting that the student model can be trained at a faster rate compared to the teacher model. For further details on the distillation procedure, please refer to [39]. The knowledge distillation procedure is illustrated in Figure 2.

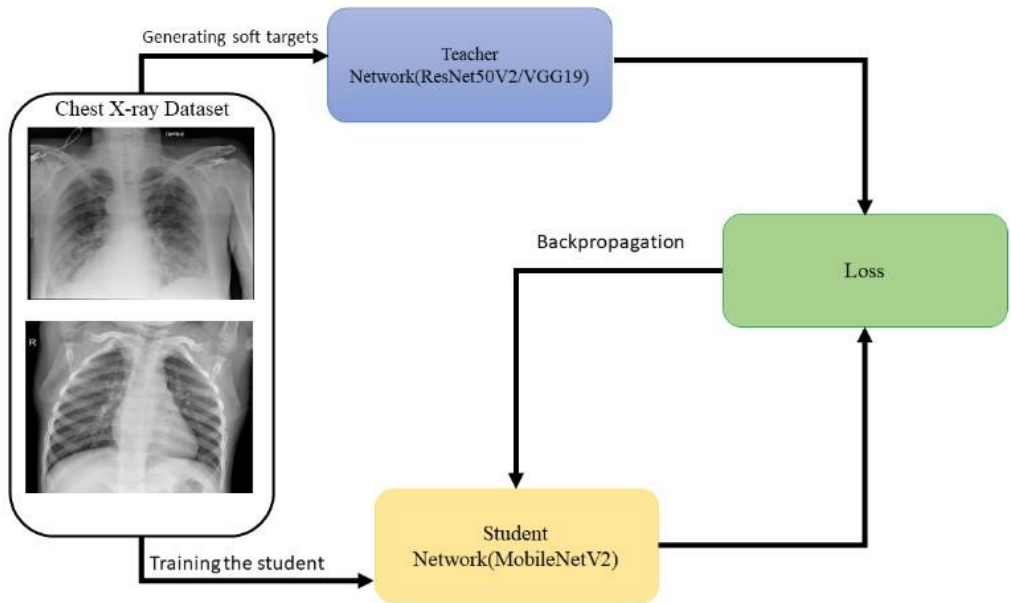

**Fig 2. Knowledge Distillation** [38] **for COVID-19 detection**

## 4. Dataset

Two public datasets were used to train the proposed deep learning model and build the required dataset. Firstly, a public dataset available at (https://github.com/ieee8023/covid-chestxray-dataset) ↗) was used for positive samples of COVID-19. Afterwards, the dataset available at (https://www.kaggle.com/c/rsna-pneumonia-detection-challenge) ↗) was utilized to collect negative samples (normal cases).

Following the merging of the two datasets, a total of 118 COVID-19 cases and 8,851 normal cases were established. It became evident that an imbalanced dataset had been created, with a significantly lower number of positive cases (COVID-19) compared to the number of normal chest X-ray medical images. To address this issue, sampling techniques were employed. The primary approach involved selecting an equal number of items from each category for the binary classification task. The oversampling method was utilized to increase the number of COVID-19 (positive cases) samples, ensuring an equal representation of both positive and negative classes.

The number of positive cases increased to 8,851 following the oversampling technique, while the number of negative (normal) samples remained unchanged. It is important to note that no pneumonia images were included in this study. Pneumonia is categorized into various classes, such as SARS, Streptococcus, ARDS, and Pneumocystis. Treating all these categories as a single class was considered impractical. Consequently, the development of a pneumonia-type classifier was postponed for future investigation. Figure 3 illustrates a subset of patients with COVID-19 and normal images obtained from the dataset.

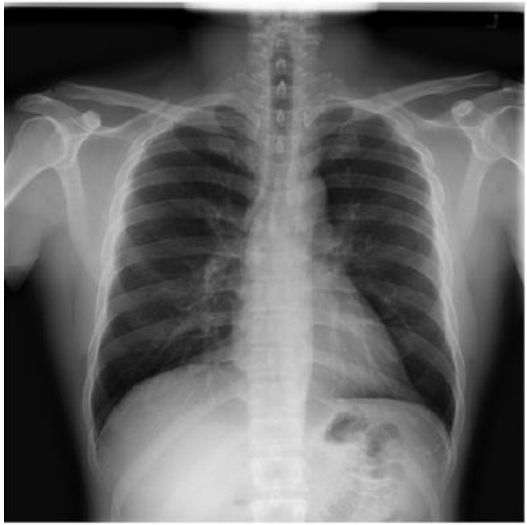

**Fig 3(a): Healthy person**

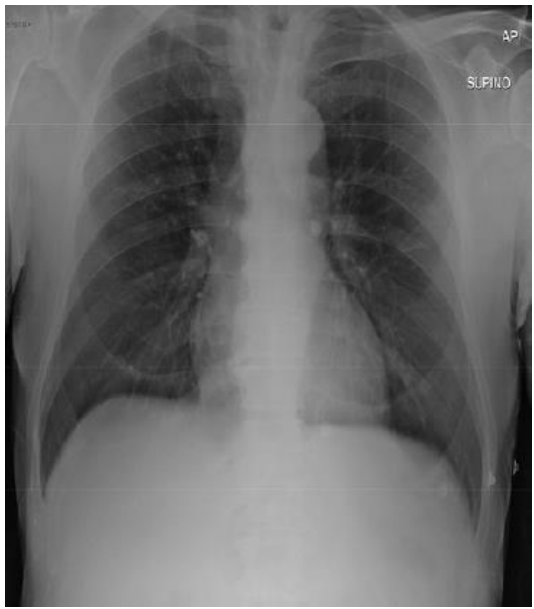

**Fig 3(b): Patient with COVID-19**

*4.1 Data Augmentation*

Prior to data augmentation, the images underwent normalization to mitigate potential problems arising from vanishing and exploding gradients. Subsequently, the images were resized to dimensions of 224x224. To improve the model's adaptability to variations present in the medical images, data augmentation techniques were applied. It was assumed that these variations did not alter the definition of the label (ground truth). Among the data augmentation approaches considered for the dataset, random rotation within the range of 0 to 20 degrees was employed. This not only helps prevent overfitting but also facilitates faster convergence of the learning curve.

5. **Training Phase**

Approximately 80% of the dataset was allocated for the training phase, while the remaining 20% was reserved for the test phase. To ensure accurate performance evaluation, the K-fold Cross-Validation technique was employed. The loss function utilized was binary Cross-Entropy, and the optimizer chosen was Adam, with a learning rate of 1e-5. A total of 5 folds (k=5) were used for Cross-Validation. The batch size was set to 32, and the model was trained for a total of 120 epochs.

*5.1 Evaluation Metrics*

To evaluate the model's performance on the test dataset, several specific criteria need to be defined. True Positive refers to correctly identifying COVID-19 positive cases among both positive and normal cases. True Negativity involves accurately identifying normal cases. False Positive occurs when COVID-19 cases are mistakenly classified as normal. False Negative refers to misclassifying normal cases as COVID-19 cases.

Precision is defined as the ratio of True Positives over the sum of True Positives and False Positives.

$$Precision = \frac{TP}{TP + FP} \qquad (1)$$

Sensitivity is the ratio of True Positives over to the sum of True Positives and False Negatives.

$$Sensitivity(recall) = \frac{TP}{TP + FN} \qquad (2)$$

Specificity is the ratio of True Negatives over the sum of False Positives and True Negatives.

$$Specificity = \frac{TN}{TN + FP} \qquad (3)$$

Finally, three additional criteria were employed to assess the performance of the proposed model on the dataset used in the study. These criteria include F1 Score, Balanced Accuracy, in addition to the conventional accuracy metric.

$$F1 - Score = \frac{2 * precision * recall}{(precision + recall)} \qquad (4)$$

$$Balanced\ Accuracy = \frac{specificity + sensitivity}{2} \qquad (5)$$

$$Accuracy = \frac{TP + TN}{Positive + Negative} \qquad (6)$$

## 6. Results and Discussion

In this section, we present the results of applying our method to the aforementioned dataset. The test images used in the evaluation were unseen by the algorithm prior to this analysis. Hence, these results demonstrate the capability of the proposed algorithm to perform effectively on previously unseen medical images.

Table 1 displays the results of the training teacher model. It is evident that the teacher model exhibits satisfactory performance in detecting chest X-ray images with good accuracy and F1-score.

Moving on to Table 2, we observe the results of the MobileNetV2 (student model). The performance of the student model in the classification task is deemed satisfactory. However, despite its favorable results, we decided to enhance its classification ability through the knowledge distillation approach.

Table 3 presents the results of the student network after completing the knowledge distillation process. As a result, the student network performs on par with existing methods in the literature and in some cases achieves even better accuracy and F1-score compared to previous publications.

Table 1: Evaluation results of the Teacher Network (ResNet50V2/VGG19)

| Fold No. | Acc | Precision | Specificity | Recall(sensitivity) | F1 score | Balanced-Acc |
|---|---|---|---|---|---|---|
| 1 | 0.978 | 0.978 | 0.978 | 0.978 | 0.978 | 0.978 |
| 2 | 0.989 | 0.978 | 1.000 | 1.000 | 0.988 | 1.000 |
| 3 | 1.000 | 1.000 | 1.000 | 1.000 | 1.000 | 1.000 |
| 4 | 0.989 | 0.978 | 1.000 | 1.000 | 0.988 | 1.000 |
| 5 | 1.000 | 1.000 | 1.000 | 1.000 | 1.000 | 1.000 |
| **Average** | **0.992** | **0.987** | **0.996** | **0.996** | **0.991** | **0.996** |

Table 2: Evaluation results of the Student Network (MobileNetV2) *Before* Knowledge-Distillation

| Fold No. | Acc | Precision | Specificity | Recall(sensitivity) | F1 score | Balanced-Acc |
|---|---|---|---|---|---|---|
| 1 | 0.987 | 0.978 | 1.000 | 1.000 | 0.992 | 1.000 |
| 2 | 0.974 | 0.976 | 0.976 | 0.976 | 0.972 | 0.976 |
| 3 | 0.974 | 0.976 | 0.976 | 0.976 | 0.972 | 0.976 |
| 4 | 0.987 | 0.978 | 1.000 | 1.000 | 0.992 | 1.000 |
| 5 | 0.987 | 0.978 | 1.000 | 1.000 | 0.992 | 1.000 |
| **Average** | **0.980** | **0.977** | **0.990** | **0.990** | **0.984** | **0.990** |

Table 3: Evaluation results of the Student Network (MobileNetV2) *After* Knowledge-Distillation

| Fold No. | Acc | Precision | Specificity | Recall(sensitivity) | F1 score | Balanced-Acc |
|---|---|---|---|---|---|---|
| 1 | 0.974 | 0.976 | 0.976 | 0.976 | 0.972 | 0.976 |
| 2 | 1.000 | 1.000 | 1.000 | 1.000 | 1.000 | 1.000 |
| 3 | 0.978 | 0.978 | 1.000 | 1.000 | 0.992 | 1.000 |
| 4 | 1.000 | 1.000 | 1.000 | 1.000 | 1.000 | 1.000 |
| 5 | 0.989 | 0.978 | 1.000 | 1.000 | 0.988 | 1.000 |
| | | | | | | |
| **Average** | **0.988** | **0.987** | **0.996** | **0.996** | **0.991** | **0.996** |
| **%Improvement** | **0.8 %** | **1.0 %** | **0.6 %** | **0.6 %** | **0.7 %** | **0.6 %** |

Table 4: Number of total parameters in each architecture

| | |
|---|---|
| **Teacher Network (ResNet50V2/VGG19)** | **49,222,390** |
| **Student Network (MobileNetV2)** | 2,334,966 |
| **Number of Parameters Reduction (%)** | **-95.3 %** |

It is important to highlight that the student network has shown superior performance compared to its previous version (without knowledge distillation) in terms of evaluation metric performance. This improvement can be attributed to the impact of knowledge distillation, which enhances the performance of MobileNetV2 to a certain extent. Additionally, knowledge distillation helps the network mitigate common neural network forgetting issues.

The performance of the student model demonstrates its suitability for medical recognition tasks in embedded systems, as it requires minimal computation due to the utilization of depthwise convolutional layers and knowledge distillation.

Table 4 illustrates the number of parameters in each architecture. Notably, knowledge distillation not only enhances the performance of the student model but also reduces the number of parameters by approximately 95.3% while maintaining performance. Therefore, the student model emerges as a viable candidate for the COVID-19 recognition task.

## 6.1 Comparison with Related articles

Conducting a comprehensive comparison with related works presents certain challenges that need to be acknowledged. Firstly, there is a significant number of chest X-ray datasets available for COVID-19 detection, with studies utilizing individual datasets or combined datasets from multiple sources. This diversity in dataset usage makes direct comparisons challenging. Secondly, some studies utilize private datasets that are not accessible to other researchers, further impeding direct comparisons. Thirdly, variations in classification tasks, such as binary or multi-class classification, add complexity to the comparison process. Additionally, the modality used, whether CT-scan images or chest X-ray images, introduces another layer of difficulty as different algorithms are specifically designed for each modality in COVID-19 recognition.

Another important point to note is that while previous related works have utilized knowledge distillation, there are significant differences in terms of the dataset, architectures, and modalities employed compared to our study. Therefore, making direct comparisons with these works is not appropriate.

Furthermore, our research focuses on minimizing computational costs and the number of parameters while maximizing evaluation metrics. We prioritize developing an efficient model rather than constructing a complex model with excessive parameters for the purpose of comparison with state-of-the-art methods. Despite these inherent challenges, we have made efforts to compare our work with relevant studies using the same dataset and methodology for training and evaluation, employing cross-validation. These comparisons are summarized in TABLE 5.

TABLE 5. Comparison against related works

| Method | Classes | Accuracy(%) | F1-Score(%) | Sensitivity(%) | Specificity(%) |
|---|---|---|---|---|---|
| **Our Work-Student Network** | Binary | 98.80 | 99.10 | 99.60 | 99.60 |
| **Our Work-Teacher Network** | Binary | 99.20 | 99.10 | 99.60 | 99.60 |
| Duran-Lopez et al. [40] | Binary | 94.4 | 93.14 | 92.53 | 96.33 |
| Khan et al. [41] | Binary | 99.0 | 98.5 | 99.3 | 98.6 |
| Sahlol et al. [42] | Binary | 98.7 | 98.21 | Not reported | Not reported |
| Al-Waisy etal. [43] | Binary | Not reported | 98.68 | 98.72 | 98.63 |
| AlMohimeed et al. [44] | Multi-Class | 98.48 | 98.48 | 98.48 | Not reported |
| Chakraborty et al. [45] | Multi-Class | 96.43 | 93.0 | 93.68 | 99.0 |
| Loey et al. [46] | Multi-Class | 96.0 | Not reported | Not reported | Not reported |
| Ieracitano et al. [27] | Binary | 80.9 | Not reported | 82.5 | 76.6 |
| Duong et al. [47] | Multi-Class | 91.93 | 93.8 | 95.2 | Not reported |
| Deb et al. [48] | Multi-Class | 85.71 | 90 | 90 | 90 |
| Soundrapandiyan et al. [49] | Multi-Class | 96.1 | Not reported | Not reported | Not reported |
| Lasker et al. [50] | Multi-Class | 97.28 | Not reported | Not reported | Not reported |
| Agrawal et al. [51] | Binary | 93.6 | 96.65 | 93.6 | 100 |

Based on the comparison presented in Table 5, our designed teacher model exhibits superior performance compared to previous works across all evaluation metrics, including accuracy, specificity, F1-score, and sensitivity. In the case of the student network after distillation, our teacher network outperforms all reported results and demonstrates state-of-the-art performance.

Taking computational costs into consideration, if there are no constraints in reducing such costs, our teacher model can be considered as a viable candidate for implementation and deployment. However, for low computational devices, our student network can be safely utilized, as it offers a good balance between performance and computational efficiency.

## 7. Conclusion

In this paper, a novel method was developed for identifying COVID-19 medical chest X-ray images. The challenge of an unbalanced dataset was addressed through the utilization of oversampling and data augmentation techniques. The proposed model's performance was evaluated using the fivefold cross-validation method to ensure robustness and accuracy.

After applying knowledge distillation, the algorithm achieved impressive results with an accuracy of 98.8% and an F1-score of 99.1%. These high-performance metrics demonstrate the efficacy of the proposed method for the COVID-19 recognition task.

However, to further enhance the algorithm, it is recommended to incorporate more diverse datasets from different countries. The inclusion of such datasets will contribute to improving the algorithm's generalization and accuracy.

For future works on medical image datasets, several avenues are suggested. These include exploring incremental learning techniques, leveraging self-supervised deep learning methods, investigating vision transformer architectures, and exploring knowledge distillation under adversarial attacks. These directions have the potential to advance the field of medical image analysis and contribute to further improvements in COVID-19 recognition algorithms.

## Declarations

### *Competing of interest:*

The corresponding author declares that there are no conflicts of interest on behalf of all authors.

### *Ethical Approval:*

This is not applicable to this research paper.

### *Funding:*

This paper was completed without receiving any external funding or financial support. The research was conducted independently and solely by the teams' own efforts and resources. While the absence of funding does not diminish the value of the work, the author acknowledges any individuals who provided assistance or support during the research process. The paper adheres to ethical standards and appropriate citation of sources. The authors are available to provide further clarification if needed.

### *Authors Contribution:*

**AmirReza BabaAhmadi:** Conceptualization, Methodology, Investigation, Writing the code - Original Draft Preparation.

**Sahar Khalafi:** Data Preprocessing, Formal Analysis and Evaluation results, Visualization, Original Draft Preparation.

**Masoud ShariatPanahi:** Supervision, Project Administration, Writing - Review & Editing.

**Mousa Ayati:** Supervision, Project Administration, Writing - Review & Editing.

All authors have read and approved the final manuscript.

*Availability of Data and Materials:*

All datasets used in this research paper are freely available online on Kaggle.